\documentclass{sn-jnl}
\usepackage{graphicx}%
\usepackage{multirow}%
\usepackage{amsmath,amssymb,amsfonts}%
\usepackage{amsthm}%
\usepackage{mathrsfs}%
\usepackage[title]{appendix}%
\usepackage{xcolor}%
\usepackage{textcomp}%
\usepackage{manyfoot}%
\usepackage{booktabs}%
\usepackage{algorithm}%
\usepackage{algorithmicx}%
\usepackage{algpseudocode}%
\usepackage{listings}%
\restylefloat{figure}
\restylefloat{table}
\begin{document}

\title[StoryGem: Voronoi treemap Approach for Semantics-Preserving Text Visualization]{StoryGem: A Voronoi treemap Approach for Semantics-Preserving Text Visualization}

\author{\fnm{Naoya} \sur{Oda}}\email{oden6680@gmail.com}

\author{\fnm{Yosuke} \sur{Onoue}}\email{onoue.yousuke@nihon-u.ac.jp}
\equalcont{These authors contributed equally to this work.}


\abstract{
    Word cloud use is a popular text visualization technique that scales font sizes based on word frequencies within a defined spatial layout.
    However, traditional word clouds disregard semantic relationships between words, arranging them without considering their meanings.
    Semantic word clouds improved on this by positioning related words in proximity; however, still struggled with efficient space use and representing frequencies through font size variations, which can be misleading because of word length differences.
    This paper proposes StoryGem, a novel text visualization approach that addresses these limitations.
    StoryGem constructs a semantic word network from input text data, performs hierarchical clustering, and displays the results in a Voronoi treemap.
    Furthermore, this paper proposes an optimization problem to maximize the font size within the regions of a Voronoi treemap.
    In StoryGem, word frequencies map to area sizes rather than font sizes, allowing flexible text sizing that maximizes use of each region's space.
    This mitigates bias from varying word lengths affecting font size perception.
    StoryGem strikes a balance between a semantic organization and spatial efficiency, combining the strengths of word clouds and treemaps.
    Through hierarchical clustering of semantic word networks, it captures word semantics and relationships.
    The Voronoi treemap layout facilitates gapless visualization, with area sizes corresponding to frequencies for clearer representation.
    User study across diverse text datasets demonstrate StoryGem's potential as an effective technique for quickly grasping textual content and semantic structures.
}




\maketitle

\section{Introduction}\label{sec1}

One of the common text visualization techniques is word clouds, which arrange words within a defined space, scaling font sizes based on word frequencies in the text \cite{viegas2008timelines,viegas2009participatory}.
This arrangement allows users to quickly grasp the overall content of a document. However, traditional word clouds place words randomly, disregarding semantic relationships between words \cite{hearst2019evaluation}.
Words are not isolated entities; they have meaningful connections that convey relationships, and random placement overlooks these semantic links \cite{schrammel2009semantically}.

To address this limitation, several visualization methods called semantic word clouds were proposed, positioning related words in close proximity to reflect their relationships \cite{wu2011semantic,5641993}.
Compared to random word placements, prior studies show that such semantic layouts are easier to comprehend due to their ability to visually convey word relationships \cite{hearst2019evaluation,schrammel2009semantically}.
However, while semantic word clouds incorporate word relationships, they face challenges with efficiently arranging words within arbitrary spatial constraints, a task more readily achieved by traditional word clouds \cite{koh2010maniwordle,jo2015wordleplus}.

Both approaches commonly represent word frequencies through font size variations \cite{viegas2009participatory}.
However, this representation can be misleading, as longer words may appear more prominent despite lower frequencies, potentially creating false impressions for users \cite{rivadeneira2007getting}.

This paper introduces StoryGem, a novel text visualization method designed to address the limitations of traditional and semantic word clouds, specifically the lack of meaningful word placements, difficulties with achieving gapless spatial layouts, and biases introduced by string length differences.
This approach facilitates gapless arrangements and allows semantically related words, identified through hierarchical clustering of a semantic word network, to be positioned in proximity.

Rather than using font sizes to represent word frequencies, StoryGem maps these frequencies to area sizes within each region. This represents a significant departure from conventional word cloud approaches, where font size directly encodes importance. While this change may initially require users to adjust their interpretation, it offers several advantages: it allows for flexible text sizing and significantly reduces the bias caused by string length variations seen in traditional approaches. In conventional word clouds, longer words often appear more prominent regardless of their actual frequency, creating potential misinterpretations. By decoupling frequency representation (area size) from readability optimization (font size), StoryGem provides a more balanced and accurate visual representation of word importance.
The problem of optimizing each word's region size is formulated as a small-scale linear programming problem, enabling efficient computation of optimal font sizes.
To evaluate the effectiveness of StoryGems as a novel text visualization method, a user study was conducted to observe the visualization results of applying several Wikipedia articles to StoryGem, followed by a comparison with existing methods.
The user study results indicate that StoryGem outperforms existing methods, particularly in tasks involving the reading of semantically related words and in overall user preference.

The main contributions of this paper to the field of information visualization are the following three points:
\begin{itemize}
  \item Proposing a text visualization technique that arranges words in arbitrary polygonal regions spaces without gaps, while considering the semantic relationships between words.
  \item Proposing a method to maximize the font size within polygonal regions for text visualization.
  \item Revealing that the proposed text visualization based on the Voronoi treemap outperforms the existing methods in reading word semantics and user preferences through the user study.
\end{itemize}

The remainder of this paper is organized as follows.
Section \ref{sec2} reviews related work on text visualization techniques, focusing on word clouds.
Section \ref{sec3} details the proposed StoryGem method, including the construction of a semantic word network, the application of the Voronoi treemap layout, and the optimization of font sizes.
Section \ref{sec4} presents the results of a user study evaluating StoryGem's effectiveness compared to existing methods.
Section \ref{sec5} discusses the implications of the study results and the potential applications of StoryGem.
Finally, Section \ref{sec6} concludes the paper and outlines future research directions.

\section{Related Work}\label{sec2}
In this section, we review several examples of text visualization techniques and related studies, with a focus on word clouds.
We highlight the current limitations faced by these approaches, positioning our research to address these issues.

\subsection{Text Visualization}\label{subsec1}
Text visualization is a technology that converts textual data into intuitive graphical representations, allowing users to visually discern patterns and trends.
With recent proliferation of digital content such as blogs, social media, emails, and electronic publications, the volume of digitized text has expanded significantly over the past few decades, increasing the relevance of text visualization techniques.
Several survey papers \cite{alharbi2018sos,alharbi2019sos,7156366} indicate various text visualization approaches exist, serving different purposes.
A common method involves using networks to represent relationships between words, revealing semantic connections.
The most prevalent approach constructs networks based on word co-occurrence patterns.
For instance, Leydesdorff\cite{leydesdorff2004university} creates a network of related words from patent application data.
PhraseNet \cite{5290726} generates networks from arbitrary text data, based on contextual word relationships, to uncover connections within documents.
Tree-based visualization creation is another network-derived technique, such as WordTree \cite{4658133}, which displays a tree structure with a specific word as the root and related words as leaves, allowing users to explore word usage patterns.

More advanced text visualizations can reveal insights beyond word frequencies, unveiling features difficult to discern from raw text.
ThemeRiver \cite{981848} is a popular time-series analysis technique that visualizes the temporal evolution of themes within large text corpora, aiming to discover prominent topics over specific periods.
More complex approaches include ThemeDelta \cite{7001093}, which reveals intricate thematic shifts, CiteRivers \cite{7192685}, enabling chronological exploration of research articles through user interactions, and RoseRiver \cite{6875938}, an interactive extension of ThemeRiver that allows users to manipulate the visualized results.
Across these text visualization methods, the common approach involves extracting words from the textual data and determining layouts based on the specific visualization purpose.
This highlights the importance of identifying appropriate visual encodings and layouts tailored to the intended analytical goals in text visualization.

\subsection{Word Cloud}\label{subsec2}
Word clouds (or tag clouds) are among the most popular text visualization techniques for textual data.
They are primarily used to visually represent the frequency of word occurrences in sentences or documents, providing a quick and intuitive way to understand which words are important or frequently used in each text.
Word clouds \cite{viegas2008timelines}, initially known as tag clouds \cite{torres2021survey}, gained popularity in the early 2000s.
In 2009, Viegas et al. introduced Wordle \cite{viegas2009participatory}, a word cloud generator that displays words of various sizes and colors, based on their frequencies in the input text.
Wordle rapidly became popular and widely adopted in various fields, such as education, research, business, and art, leading to the widespread use of word cloud visualizations.
Word clouds continue to evolve, providing better visualization results according to user needs.
Interactive method like ManiWordle \cite{koh2010maniwordle}, Wordleplus \cite{jo2015wordleplus}, and EdWordle \cite{wang2020recloud} allow users to manipulate word cloud layouts, suggesting that customizable visualizations can align better with user objectives.
Several novel text visualization methods have been proposed by integrating word clouds with other techniques.
For instance, Xu et al. combine word clouds with ThemeRiver layouts for quick topic analysis on social media \cite{6634134}.
TreeCloud \cite{gambette2010visualising} incorporates network layouts into word clouds, whereas SentenTree \cite{7536200} improves WordTree and word cloud designs to better convey word relationships within sentences.
RadCloud \cite{6902889} applies the RadVis layout \cite{663916,5190784}, originally developed for non-textual data, to word clouds, for creating a novel text visualization approach.
These examples demonstrate the potential of combining word clouds with diverse visualization methods, both textual and non-textual, to generate readable and insightful visualizations tailored to user needs.

Although word cloud use is a popular text visualization technique, it has several limitations.
The normal word cloud layouts arrange words alphabetically or in descending order of frequency, scaling proportionally to their occurrences.
However, such basic layouts can be difficult for users to comprehend.
To improve readability, researchers have proposed layouts that arrange words more than the normal methods.
For example, Kaser and Lemire introduced an optimization technique that reduces overall white space by using the smallest possible rectangle around each word.
Other approaches leverage natural language processing, such as PrefixTagClouds \cite{6676541}, which groups words based on prefixes and color-codes them for better interpretability.
These efforts highlight the significant impact of layout and natural language processing on text data analysis tasks, motivating the development of additional readable word cloud visualizations.
Overall, word clouds continue to evolve through layout improvements, integration with other visualization techniques, and requirements for specific applications like social media analysis.
However, their inherent limitations such as disregarding semantic relationships between words and representing frequencies solely through font sizes remain to be addressed.
Overcoming these challenges can lead to more effective and insightful text visualization tools.

\subsection{Semantic Word Cloud}\label{subsec3}
While word clouds effectively visualize word frequencies, traditional layouts disregard semantic relationships between words, making it challenging to grasp the underlying context and meaning of the text.
To address this limitation, several text visualization methods, collectively referred to as semantic word clouds, were proposed.
These approaches aim to reflect the semantic connections between words in their layouts, which has shown to improve overall topic comprehension compared to random word placements \cite{hearst2019evaluation,schrammel2009semantically}.
One of the earliest semantic word cloud techniques used multidimensional scaling (MDS) to position semantically similar words in proximity, mapping their relationships onto a two-dimensional plane.
Various extensions have built upon this MDS-based word-positioning approach, such as methods for space-efficient layouts using seam carving \cite{wu2011semantic} and techniques for removing white space while preserving relative word distances \cite{5641993}.
These efforts primarily focused on creating more compact semantic word cloud layouts.
With the rise in word-embedding techniques, researchers have proposed methods that measure word similarity based on vector representations and determine layouts from similarity networks.
Recent developments include interactive systems like Edwordle \cite{wang2017edwordle} and MySemCloud \cite{huber2023mysemcloud}, which allow users to freely edit word cloud layouts while preserving semantic relationships between words.
Regardless of the specific technique, semantic word clouds aim to improve upon traditional word cloud visualizations by incorporating meaningful word arrangements that reflect their semantic connections, enabling users to better comprehend the underlying textual content and themes.
Semantic word cloud has the advantage of reflecting the semantic relationship of words in the layout, but at the expense of the compactness of a traditional word cloud.

\subsection{Text Visualization Issues}\label{subsec4}
In most text visualization approaches, individual words are extracted from the input data, and word occurrence frequencies are primarily used to determine the final visualizations.
Consequently, the formatting and preprocessing of textual data significantly impact the resulting visual output.
Additionally, the inherent length of words can substantially affect visualization outcomes, as font sizes are commonly scaled according to word importance or frequency.
This practice can lead to visual distortion, where longer words appear more prominent regardless of their actual relevance, potentially resulting in misinterpretations.
Word clouds, one of the most popular text visualization methods, have seen advancements through approaches like semantic word clouds, which aim to reveal semantic connections between words.
However, these methods often struggle to retain the efficiency of traditional word clouds, which effectively arrange words within a confined space.
In light of these challenges, this paper proposes a novel text visualization method that combines the strengths of traditional word clouds and semantic word clouds.
Our approach facilitates efficient word placement within a defined area, similar to word clouds, while enhancing the interpretability of semantic relationships between words, as seen in semantic word clouds.
By addressing the limitations of existing techniques, our method aims to provide more effective and insightful text visualizations, allowing users to interpret data with greater accuracy and clarity.

\section{Story Gem}\label{sec3}
\begin{figure*}[t]
  \begin{center}
    \includegraphics[width=\linewidth]{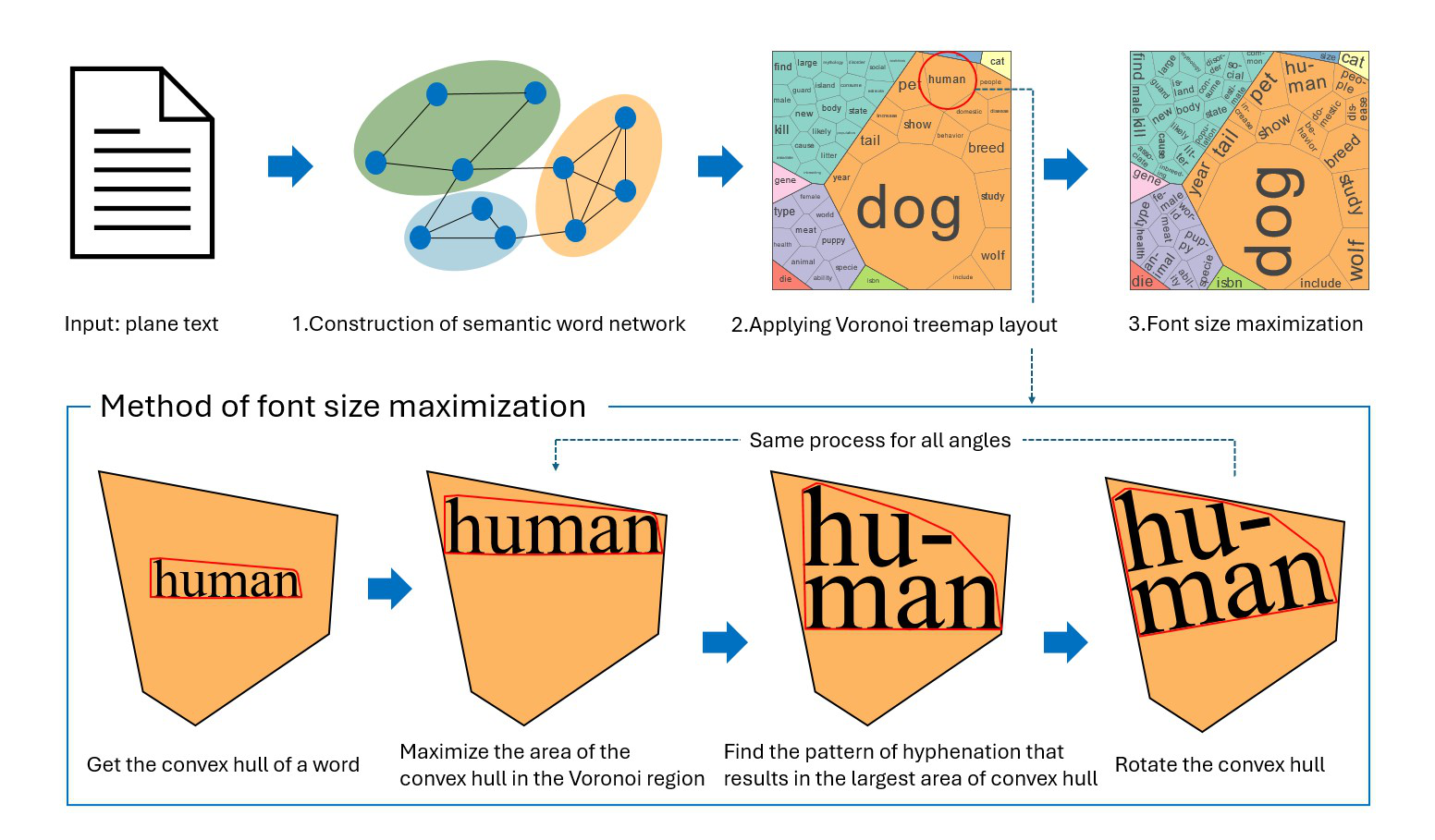}
    \caption{
      Procedure of StoryGem
    }\label{methods}
  \end{center}
\end{figure*}
In this section, we detail our proposed text visualization method.
The primary goal, akin to word clouds, is to enable quick comprehension of entire documents and identify frequently used words.
However, our approach aims to preserve word semantics and clearly convey the overarching themes.
As Fig. \ref{methods}  shows, there are three main steps in this method.
\begin{enumerate}
  \item Construction of a semantic word network
  \item Applying the Voronoi treemap layout
  \item Maximizing the font size
\end{enumerate}
By following this systematic process, our technique facilitates efficient spatial word arrangements while emphasizing semantic relationships between words, ultimately providing more insightful visualizations compared to traditional word cloud approaches.

\subsection{Construction of Semantic Word Network}\label{subsec5}
The proposed method accepts plain text as input data.
There are some words such as ``it'', ``is'', and ``a'' that appear frequently in general sentences; however, they do not have a significant meaning.
These words are removed as stop words during natural language processing, because they can negatively affect the visualization results.
In addition to the standard stop words in English natural language processing obtained from the Natural Language Toolkit (NLTK), symbols and numbers are applied as stop words when English text is input to the proposed method.
As verbs, nouns, and adjectives are parts of speech that have strong meanings, we identify the part of speech of each word and consider words that are not verbs, nouns, or adjectives to be stop words.
Through this process, a set of words $W$ is obtained from the input plain text data after removing the stop words.
As the number of times a word appears in a document is one of the important factors indicating the importance of the word, we examine the number of times that the words $w_1, w_2, \ldots, w_n (i = 1,\cdots,n)$ in the set $W$ appear in the original text data.

Next, we extract the word vector (a numerical representation of word meaning in high-dimensional space, where words with similar meanings have similar vector representations) of each word $w_1, w_2, \ldots, w_n (i = 1,\cdots,n)$, which is necessary to obtain similarity between words.
To obtain word vectors in this paper, we use a pretrained model called FASTTEXT\cite{grave2018learning}, which is available in the natural language processing library FASTTEXT \cite{joulin2016bag}.
We use this pretrained model because it has word vectors for 157 languages, learned from CommonCrawl and Wikipedia using FastText, making it scalable for many languages.
When extracting word vectors, we remove words that are not included in this training model because it is difficult to extract vectors under the same conditions as other words.
By performing these processes, we obtain from the input plain text data the number of occurrences in the text and the word vectors of words that are not included in the stop words and contained in the learned model.

Next, a k-nearest neighbor graph (a graph structure where each node is connected to its k most similar neighbors, where k is a user-defined parameter) is constructed based on the similarity between words, with each word obtained in the preprocessing stage as a node.
The similarity between words is calculated from the word vector of each word, using cosine similarity (a measure of similarity between two vectors that calculates the cosine of the angle between them, with values closer to 1 indicating greater similarity).
For each word, the k-nearest neighbor graph is created by stretching the edges by $k \in \mathbb{N}$, starting from the word with the largest cosine similarity value.
\subsection{Applying Voronoi Treemap}\label{subsec6}
Clustering is performed on the obtained k-nearest neighbor graph to separate highly related words into several groups.
Considering the hierarchical clustering based on the characteristics of VoronoiTreemap\cite{balzer2005voronoi} to be used later and the computational time and clustering accuracy, we use the Louvain method \cite{blondel2008fast}, which is a type of modularity clustering that efficiently detects communities in networks by optimizing the modularity quality function.
The Louvain method has randomness at the initialization stage of clustering and thus shows the property that results are different for various runs.
By fixing the random seed value, we ensure reproducibility by obtaining the same results across various runs.
By constructing and clustering these k-nearest neighbor graphs, we obtain hierarchically clustered data based on the list of words extracted from input data.

The data obtained in the process up to this point are applied to VoronoiTreemap layout \cite{balzer2005voronoi}.
VoronoiTreemap leverages centroidal Voronoi tessellations (CVT) to generate hierarchical layouts that are gapless and non-overlapping.
The algorithm starts by initializing a bounded polygonal space, representing the root node of the hierarchy, along with a set of points (called generators) and weights corresponding to the data items.
The weights reflect the relative sizes or importance of the hierarchical nodes.
The layout process is iterative and consists of three main steps.
First, a weighted Voronoi tessellation is computed, partitioning the space into regions where each region is associated with a generator.
The weighted centroidal Voronoi tessellation ensures that regions with larger weights occupy proportionally larger areas.
Second, the weights of the generators are adjusted iteratively to match the desired areas of the regions with the corresponding data sizes.
If a region is smaller than its target size, the weight of the generator is increased, and the tessellation is recalculated. Third, each generator is moved to the centroid (center of mass) of its corresponding region, reducing spatial imbalance and improving layout accuracy.

Unlike traditional Treemaps, which are limited to rectangular layouts, VoronoiTreemap uses polygonal regions, allowing the layout to adapt to arbitrary shapes such as circles or irregular polygons.
Furthermore, its use of centroidal Voronoi tessellations minimizes aspect ratio distortions and enhances the interpretability of hierarchical relationships.
We focused on the potential of VoronoiTreemap to address the challenges faced by text visualization methods due to its gapless space partitioning and ability to represent data importance through area allocation.
\subsection{Font Size Maximization}\label{subsec7}
Adjust the font size to draw the word as large as possible within each Voronoi area to increase the visibility of the word.
Since considering the word's drawing area as it is would be complicated, consider the word's convex hull as the word's drawing area.
The area drawn using VoronoiTreemap is guaranteed to be a convex polygon.
In other words, maximizing the area of a word to be drawn within each Voronoi region can be replaced by the problem of maximizing the area of another convex polygon to be placed within any given convex polygon.
This allows us to maximize the area of the convex polygons (word convex hulls) placed inside the outer convex Voronoi regions, achieving our goal of maximizing word visibility within the allocated space.

Previous research\cite{agarwal1998largest} by Agarwal et al. proposed an algorithm for placing one convex polygon inside another by rotating and scaling it to achieve the maximum possible area.
Chan et al. extend the algorithm proposed by Agarwal et al. to reduce the computational complexity \cite{chan2024convex}.
However, their study focused on convex polygons in general and did not specifically address word placement.
As a result, their approach imposes no restrictions on the rotation of the inner convex polygon.
When applying their method to word placement, there is a risk that words could be rotated up to 180°, significantly impairing readability.
To address this limitation, our study introduces a method that restricts the rotation of each word (represented as a convex polygon) to ensure it maintains readability while still maximizing its area within the corresponding Voronoi region.

Let $Q$ be the convex hull of the word whose area is maximized and $O$ be the convex hull of the original word.
By scaling the polygon $O$ by a factor of $S$ and translating it by $d_x, d_y$ along the x-axis and the y-axis, respectively, we obtain $Q$ such that $O$ overlaps $Q$.
This affine transformation is described by the following formula, where $x^o_i, y^o_i, x^q_i, y^q_i$ denote the $x, y$ coordinates of any vertex of the convex polygons $O, Q$.
The scaling and translation of a convex polygon can be expressed as an affine transformation as in Equation \ref{eq:1}, and this equation can be rearranged for $x_i^o,y_i^o$ to obtain Equation \ref{eq:2},\ref{eq:3}.
\begin{align}\label{eq:1}
  \begin{pmatrix}
    x^o_i \\
    y^o_i \\
    1     \\
  \end{pmatrix}
  =
  \begin{pmatrix}
    1 & 0 & d_x \\
    0 & 1 & d_y \\
    0 & 0 & 1   \\
  \end{pmatrix}
  \begin{pmatrix}
    S & 0 & 0 \\
    0 & S & 0 \\
    0 & 0 & 1 \\
  \end{pmatrix}
  \begin{pmatrix}
    x^q_i \\
    y^q_i \\
    1     \\
  \end{pmatrix}
\end{align}
\begin{align}
  x_i^o = Sx_i^q + d_x \label{eq:2}\\
  y_i^o = Sy_i^q + d_y \label{eq:3}
\end{align}
Organize Equations \ref{eq:2} and \ref{eq:3} for $d_x,d_y$.
In this case, the translation values $d_x,d_y$ in the $x$-axis and $y$-axis directions consider the same value for any vertex considered, so $i=1$.
Also, the scaling factor $S$ of the reduction follows from its definition.
\begin{align*}
  d_x = x_1^o - Sx_1^q \\
  d_y = y_1^o - Sy_1^q \\
  S = \frac {x^o_2 - x^o_1}{x^q_2 - x^q_1}
\end{align*}
By substituting the values of $S,d_x,d_y$ into the Equations \ref{eq:1} and \ref{eq:2} and rearranging with respect to $q$, Equations \ref{eq:4} and \ref{eq:5} are obtained.
\begin{align}
  &(x_2^o - x_1^o) x_i^q - (x_2^o - x_i^o) x_1^q + (x_1^o - x_i^o) x_2^q                          = 0 \label{eq:4}\\
  &(x_2^o - x_1^o) y_i^q - (x_2^o - x_1^o) y_1^q + (y_1^o - y_i^o) x_2^q + (y_i^o - y_1^o) x_1^q  = 0 \label{eq:5}
\end{align}
Next, consider that each vertex $q_1, q_2, \dots ,q_m$ of the optimized polygon $Q$ all have coordinates inside the outer polygon $P$.
These can be rephrased as the coordinates $x^q_j,y^q_j$ of any vertex of the optimized polygon $Q$ being the convex combination of each vertex $(x^p_1,y^p_1), \dots ,(x^p_n,y^p_n)$ of the outer polygon.
The fact that any vertex $x^q_j$ of the optimized polygon $Q$ is a Thus, the coordinates of each vertex $(x^q_i,y^q_i)$ of the optimized polygon $Q$ can be expressed as $x^q_i = \sum\limits_j \lambda_{ij}x^p_j, y^q_i = \sum\limits_j \lambda_{ij}y^p_j$, where $(x^p_1,y^p_1), \dots ,(x^p_n,y^p_n)$ are the vertices of the outer polygon $P$ and $(x^o_1,y^o_1), \dots ,(x^o_m,y^o_m)$ are the vertices of the inner polygon $O$.
can be expressed by the following equation using real numbers $\lambda _i$ that satisfy $0 \leq \lambda _i \leq 1$ and $\sum\limits _{i=1}^n \lambda _i = 1$.
\begin{align}\label{eq:6}
  q_j = \lambda _1 p_1 + \lambda _2 p_2 + \lambda _3 p_3 + \dots + \lambda _n p_n
\end{align}
By considering Equation \ref{eq:6} and organizing the values for $q$ in Equations \ref{eq:4} and \ref{eq:5} we obtain the following equations.
\begin{equation}
  \begin{aligned}
    & (x^o_2 - x^o_1) \sum_{j} \lambda _{ij}x^p_j 
    - (x^o_2 - x^o_i) \sum_{j} \lambda _{1j}x^p_j 
    + (x^o_1 - x^o_i) \sum_{j} \lambda _{2j}x^p_j 
    = 0 \label{eq:7}
  \end{aligned}
\end{equation}
\begin{equation}
  \begin{aligned}
    & (x^o_2 - x^o_1) \sum_{j} \lambda _{ij}y^p_j 
    && - (x^o_2 - x^o_i) \sum_{j} \lambda _{ij}y^p_j \\
    &  && + (y^o_1 - y^o_i) \sum_{j} \lambda _{2j}x^p_j 
    + (y^o_i - y^o_1) \sum_{j} \lambda _{1j}x^p_j 
    = 0 \label{eq:8}
  \end{aligned}
\end{equation}  
To maximize the font size, we can maximize the area of the convex hull of a word, which is equivalent to maximizing the distance $x_j^q-x_{j-1}^q$ along the $x$ axis between any two points of the convex hull.
Assuming $j=2$, let $z = x_2^q-x_1^q$ and consider maximizing this value.
This value of $z$ is expressed as in Equation \ref{eq:9} using Equation \ref{eq:6}.
\begin{align}\label{eq:9}
  z = \sum\limits _j \lambda _{2j} x^p_j  - \sum\limits _j \lambda _{1j} x^p_j
\end{align}
From the above results, we can maximize the value of $z$ while satisfying the constraints $0 \leq \lambda_i \leq 1$ and $\sum_{i=1}^n \lambda_i = 1$ in Equations \ref{eq:7} and \ref{eq:8} and $\lambda_{ij}$ to maximize the font size.
As this constraint and the value to be maximized are all linear expressions, the requirement is satisfied by solving the following linear programming problem with $z$ as the objective function and other expressions as constraints:
\begin{equation*}
  \begin{aligned}
     & \text{maximize}   &  & \sum_j \lambda _{2j} x^p_j  - \sum_j \lambda _{1j} x^p_j                                                                      \\
     & \text{subject to} &  & (x^o_2 - x^o_1) \sum_{j} \lambda _{ij}x^p_j - (x^o_2 - x^o_i) \sum_{j} \lambda _{j}x^p_j                                     \\
     &                   &  & \quad + (x^o_1 - x^o_i) \sum_{j} \lambda _{2j}x^p_j = 0,                                             &  &  & 1 \leq i \leq n, \\
     &                   &  & (x^o_2 - x^o_1) \sum_{j} \lambda _{ij}y^p_j - (x^o_2 - x^o_1) \sum_{j} \lambda _{1j}y^p_j                                     \\
     &                   &  & \quad + (y^o_1 - y^o_i) \sum_{j} \lambda _{2j}x^p_j + (y^o_i - y^o_1) \sum_{j} \lambda _{1j}x^p_j= 0,&  &  & 1 \leq i \leq n, \\
     &                   &  & \sum_j \lambda _{ij} = 1,                                                                            &  &  & 1 \leq i \leq n, \\
     &                   &  & 0 \leq \lambda_{ij} \leq 1,\quad 1 \leq i \leq n,1 \leq j \leq m.                                    &  &  &
  \end{aligned}
\end{equation*}

The linear programming problem above has a clear geometric interpretation. The objective function $\sum_j \lambda_{2j} x^p_j - \sum_j \lambda_{1j} x^p_j$ represents the horizontal distance between two vertices of the word's convex hull after transformation, which directly correlates with the word's size. Maximizing this value effectively maximizes the font size.

The first set of constraints $(x^o_2 - x^o_1) \sum_{j} \lambda_{ij}x^p_j - (x^o_2 - x^o_i) \sum_{j} \lambda_{1j}x^p_j + (x^o_1 - x^o_i) \sum_{j} \lambda_{2j}x^p_j = 0$ ensures that the shape of the word is preserved during transformation. These equations maintain the relative positions of all vertices in the word's convex hull, preventing distortion of the word's appearance.

The second set of constraints $(x^o_2 - x^o_1) \sum_{j} \lambda_{ij}y^p_j - (x^o_2 - x^o_1) \sum_{j} \lambda_{1j}y^p_j + (y^o_1 - y^o_i) \sum_{j} \lambda_{2j}x^p_j + (y^o_i - y^o_1) \sum_{j} \lambda_{1j}x^p_j= 0$ similarly preserves the vertical relationships between vertices.

The constraint $\sum_j \lambda_{ij} = 1$ ensures that each vertex of the transformed word is expressed as a convex combination of the vertices of the Voronoi region, meaning it remains within the region's boundaries. Finally, $0 \leq \lambda_{ij} \leq 1$ guarantees that the convex combination is valid, keeping the transformed word entirely contained within the Voronoi region.

By solving this linear programming problem for each word, we determine the optimal placement and scaling that maximizes the word's size while ensuring it fits completely within its designated Voronoi region, maintaining readability and preserving the semantic relationships represented by the region's position and area.

Assuming the case where the visibility of long strings of English words is significantly impaired, we attempt to adjust the length of word strings by hyphenating the words.
We search for syllable breaks in the word and consider them as positions that can be hyphenated.
When there are $n$ possible insertion positions, we solve the optimization problem for a total of $2^n$ patterns with and without hyphen insertion for all positions, and adopt the hyphen insertion pattern with the largest scaling factor.

We can also rotate the word and update the coordinates and then solve a similar linear programming problem to determine the largest font size in the Voronoi region.
However, if the word is rotated from its normal angle to nearly 180 degrees, as in a conventional word cloud, the word becomes difficult to read and may significantly affect the visibility of the visualization results.
To address this, we constrain the rotation angle $\theta$ of a word from its initial placement to be within the range $-90^\circ \leq \theta \leq 90^\circ$.
Theoretically, it is possible to maximize the font size within a region by rotating the word by very small angular intervals within this $\theta$ range and finding the optimal font size for each angle.
However, this process is computationally expensive for every word.
The computation time can be reduced by increasing the interval of rotation angles, so the width of the rotation angle is determined such that the computation time is acceptable for the user to obtain the visualization results.

\subsection{Implementation Example}\label{subsec8}
\begin{figure}[H]
  \centering
    \includegraphics[width=\linewidth]{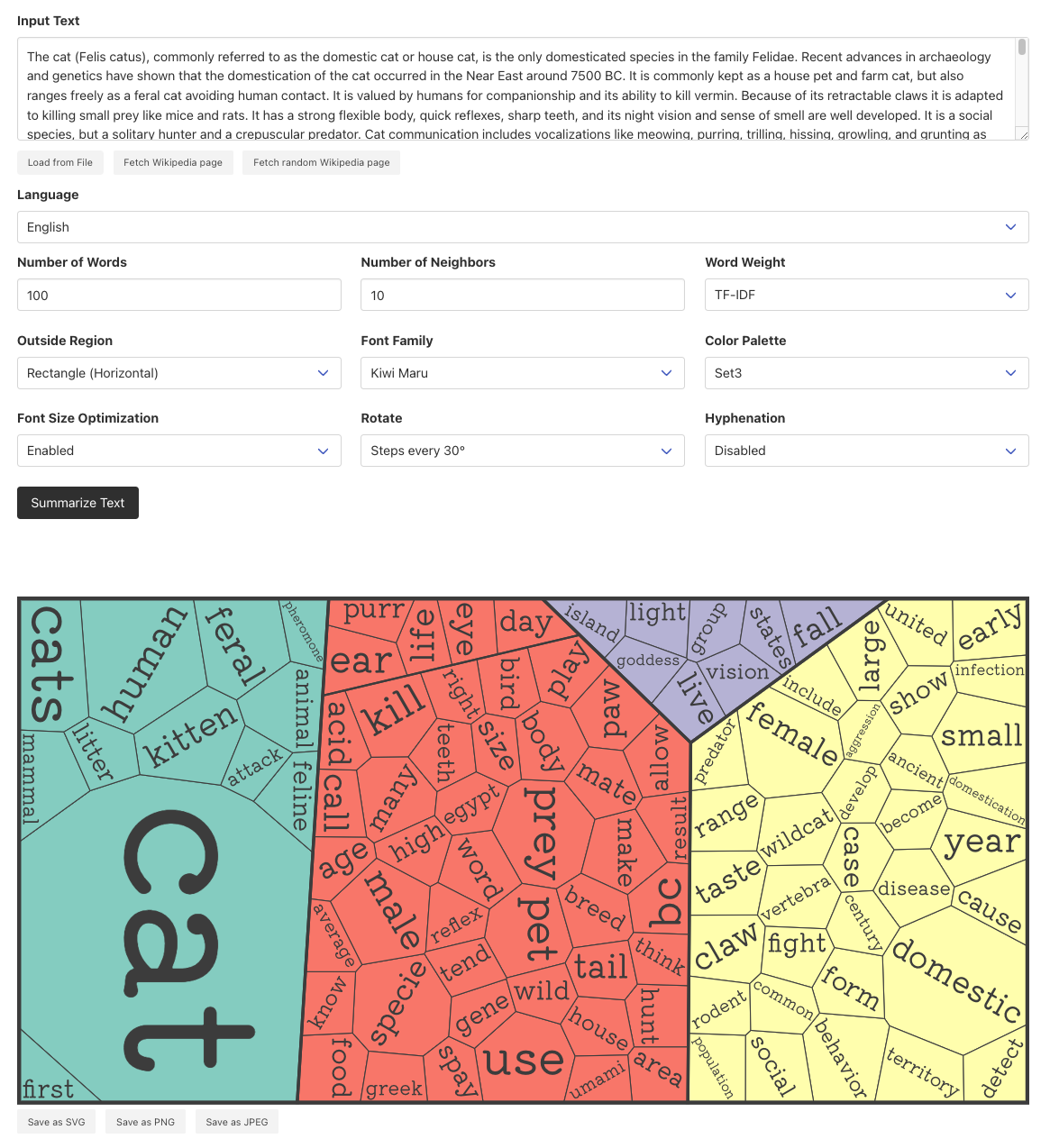}
  \caption{Screenshot of the implementation example of StoryGem:}\label{fig:system-overview}
\end{figure}

We implemented the proposed method as a web application (\url{https://storygem.vdslab.jp/}) that users can try freely.
Fig. \ref{fig:system-overview} shows the overall view of the implemented web application.
In this application, the user can freely set 10 parameters, including arbitrary text data, language, number of words, k-value of k-nearest neighbor graph, weighting of words, drawing space, font, font size optimization, sense of word rotation during optimization, and hyphenation.

\section{Application Examples}\label{sec4}

\begin{figure*}[t]
  \begin{center}
    \includegraphics[width=\linewidth]{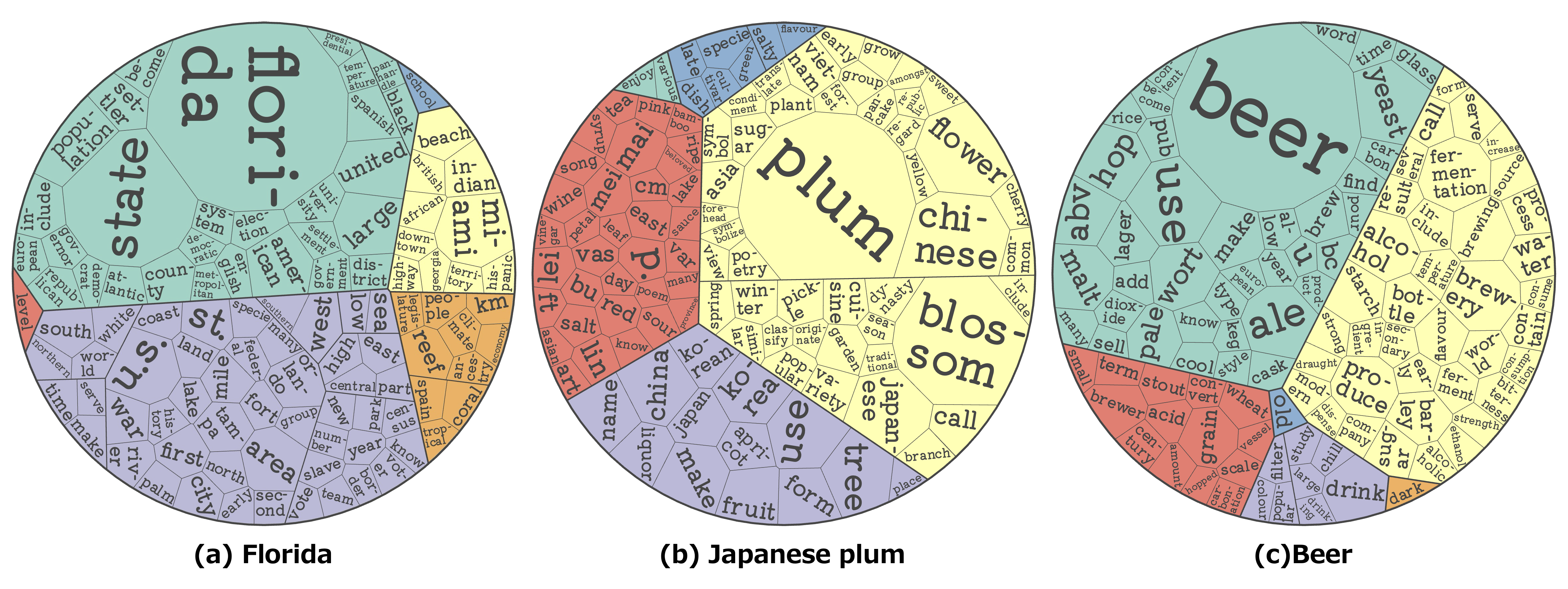}
    \caption{Visualization results of the Wikipedia dataset using StoryGem. (a) Florida, (b) Japan Plum, and (c) Beer.}\label{fig:Wikipedia}
  \end{center}
\end{figure*}

We provide an example using the Wikipedia dataset as experimental data.
Wikipedia pages are commonly used in text visualization techniques, such as word clouds, to illustrate practical applications.
In this study, we also used Wikipedia texts for our application examples.
The examples included texts from four categories—people, animals, sports, and food—sourced from English Wikipedia articles.
Here, we summarize additional application examples that were not included in the evaluation experiments, applying StoryGem to articles on topics outside these categories: Japanese plum (plants), Beer (beverages), and Florida (places).

The visualization results in Fig. \ref{fig:Wikipedia} provides an intuitive grasp of the key features of each topic.
In (a), the prominent words ``florida'' and ``state'' immediately indicate that the visualization pertains to the state of Florida.
In StoryGem, a word's importance is visually represented by its allocated area, with the most important words occupying the largest regions.
This feature enhances the prominence of high-importance words compared to conventional text visualizations. The yellow and purple regions highlight words related to the ocean and coast, such as ``beach'', ``palm'', and ``coast'', signifying the significance of Florida's beaches and shores as tourist attractions.
Major cities like ``miami'', ``orlando'', and ``tampa'', also stand out, emphasizing their importance.

In (b), words associated with flowers and blooming—such as ``flower'', ``blossom'', and ``plum'' are most prominent, suggesting that the topic involves plum blossoms.
StoryGem clusters related words within similarly colored areas, facilitating intuitive and quick comprehension of trends within the text data. For instance, the purple area includes names of East Asian countries, such as ``japan'', ``china'', and ``korea'', indicating that plum blossoms are primarily found in East Asia.

In (c), words describing beer as an alcoholic beverage, including ``drink'', ``alcohol'', and ``beer'', are emphasized, making it clear that the information pertains to beer.
StoryGem not only highlights high-importance words but also enhances the visibility of less prominent terms, offering a more balanced view of the content.
The green area contains words related to ingredients, such as ``hop'', ``malt'', and ``yeast'', enabling viewers to infer aspects of the brewing process from the visualization alone.

This example demonstrates that StoryGem is an effective tool for extracting essential information from large volumes of text data and presenting it in an intuitive, easily comprehensible format.
StoryGems unique approach to spatially organizing words by importance and semantic grouping allows users to quickly gain insights and interpret key topics within the data.

\section{User Study}\label{sec5}
We aim to evaluate the superiority of our proposed method over conventional text visualization techniques across three key aspects: recognition time, accuracy, and user satisfaction.
Additionally, we explore the impact of optimizing font size on these aspects within our proposed method.
To achieve these objectives, participants are tasked with reading visualizations generated from four types of text data using four various text visualization methods, including our proposed approach.
They are then measured on the time taken to complete the task and asked to respond to simple questions related to frequency search, relationship search, and document estimation tasks.
Following each visualization, participants undergo a mental workload assessment using the NASA Task Load Index (NASA-TLX).
Finally, participants provide subjective preferences for the four visualization methods and are invited to freely interact with the system using our proposed method, offering feedback.
Quantitative analysis is conducted based on task completion times and the accuracy of responses to questions, while qualitative analysis relies on user feedback collected during the interactive session.
This comprehensive evaluation approach allows us to assess the effectiveness and usability of our proposed text visualization technique in comparison to existing methods.
\begin{figure*}[t]
  \begin{center}
    \includegraphics[width=\linewidth]{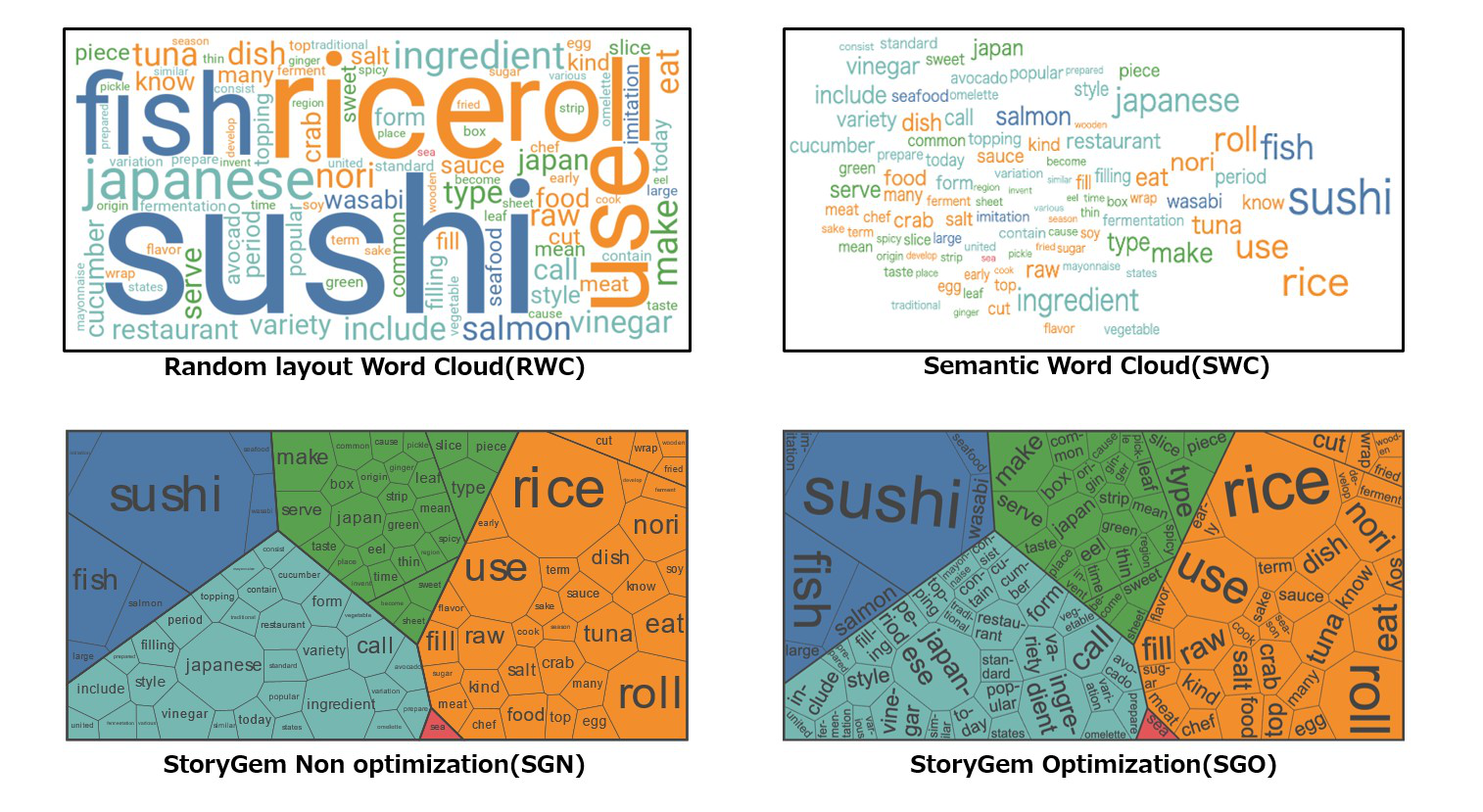}
    \caption{Four text visualization examples used in the user study, applied to the ``Sushi'' page from Wikipedia. In StoryGem (SGN and SGO), the color assigned to the drawing area of each word represents its cluster, whereas in RWC and SWC, the color of the word itself corresponds to the cluster. The size of the drawing area of a word in StoryGem represents its importance, whereas in RWC and SWC, the size of the word itself indicates its importance. }\label{fig:evosample}
  \end{center}
\end{figure*}

\subsection{Experimental Conditions}\label{subsec9}
The user study consists of the following six steps:
\begin{enumerate}
  \item Input subject information
  \item Explanation of how to read the visualization results
  \item Task evaluation experiment
  \item Mental workload evaluation using NASA-TLX
  \item Ranking preferences of visualization methods
  \item System use and feedback on StoryGem
\end{enumerate}
Considering the potential influence of individual personality traits on visualization reading, we collect non-identifiable subject information, including age, gender, and knowledge of information visualization.
We provide detailed instructions on how to read and interpret the features of three text visualization types: word cloud, semantic word cloud, and StoryGem.
Users practice reading each visualization type, ensuring comprehension before proceeding.

\subsubsection{Participant Demographics}
For this study, we recruited 22 participants from undergraduate and graduate programs in information science. All participants had at least basic knowledge of data visualization concepts, though their experience levels varied. Prior to the experiment, we collected demographic information including age (ranging from 21-25 years, mean=21), gender distribution (19 male, 3 female), academic background, and prior experience with visualization techniques. Sixteen participants were currently majoring in information visualization, while six had more limited exposure to the field. While all participants reported familiarity with the concept of word clouds, none had extensive prior experience creating or analyzing them professionally, and eight had never heard of semantic word clouds before the study. This demographic profile allowed us to assess the visualization methods with users who had sufficient technical background to understand the concepts, but without specialized expertise that might bias the results.

\subsubsection{Experimental Tasks}
Our experiment focused on two main timed tasks, followed by a simple comprehension assessment:

1. \textbf{Importance reading task}: This task was designed to evaluate how effectively each visualization method conveyed word frequency information. From the visualization, participants were asked to select one word they believed appeared most frequently and one word they believed appeared least frequently in the original text. We measured both task completion time and accuracy by comparing their selections against the actual word frequencies in the text. This task was crucial for assessing whether StoryGem's novel approach of using area size instead of font size could effectively communicate word importance.

2. \textbf{Word relationship reading task}: This task aimed to evaluate how well each visualization method conveyed semantic relationships between words. Participants were presented with three randomly selected target words and asked to identify one word they believed was most semantically related to each target word. We measured both completion time and accuracy, with responses considered correct if the selected word was among the top three words with highest cosine similarity to the target word (calculated using word vectors) among all displayed words. This task was particularly important for assessing whether StoryGem's clustering and layout approach improved the identification of semantic relationships compared to traditional methods.

Following these timed tasks, participants completed a simple comprehension assessment where they were asked to identify the Wikipedia article topic based solely on the visualization. This final assessment helped evaluate how effectively each method conveyed the overall theme of the text, though it was not included in the time measurements.

The evaluation comprises four visualization methods: Random Layout word cloud (RWC), semantic word cloud (SWC), StoryGem without optimized font size (SGN), and StoryGem with an optimized font size (SGO).
Participants complete tasks such as frequency and relationship searches, as well as document estimation, using these methods.
Tasks involve identifying specific features within visualization results and recording the time taken for certain tasks.
The experiment employs Wikipedia's English text data covering themes like sushi, Oda Nobunaga, soccer, and monkeys.
Participants read visualization results in a predefined order to minimize potential task repetition effects.
To mitigate sequence bias, participants are divided into four groups as shown in Table \ref{table:experiment-group}, each assigned a different order of data and visualization methods.
This randomized allocation ensures an unbiased assessment across all experimental conditions.
After completing the tasks, participants undergo a mental workload evaluation, using NASA-TLX.
They then rank their preferences for the four visualization methods and are invited to freely interact with the StoryGem system, providing feedback on its usability and effectiveness.
The visualization results used in the task evaluation experiments are unified across all visualization methods using the similarity between words, color assignment, and word weights when obtaining visualization results in StoryGem.

In StoryGem, the area of each Voronoi region represents the frequency of occurrence of a word, while in other visualization methods, the size of a word represents its frequency.
The color of a Voronoi region in StoryGem is assigned to the color of the corresponding word in other visualization methods.
For the word cloud visualization, we use Python's word cloud library to produce results with randomized word placement.
The Semantic word cloud employs the simplest layout algorithm, which transforms word vectors into coordinates in 2D Euclidean space, using Multidimensional Scaling (MDS) and removes white space using a force-directed algorithm.
In StoryGem without font size optimization, the size is adjusted so that the smallest rectangle surrounding a word is inscribed at the center of the largest circle that does not extend beyond the region centered at the center of gravity of each Voronoi region.
In StoryGem with an optimized font size, hyphenation is applied, and the word is rotated by 3 degrees steps and optimized at each angle to determine the size that is the largest within the region.
By unifying these factors across visualization methods, we ensure a consistent basis for comparison in the task evaluation experiments, allowing for a fair assessment of the effectiveness and strengths of each approach.
Participants evaluate each visualization outcome using a 10-point Likert scale questionnaire based on the NASA-TLX\cite{hart1988development}.
\begin{table}[H]
  \centering
  \begin{tabular}{|c|c|c|c|c|}
    \hline
      & Sushi & Oda & Football & Monkey \\ \hline
    A & RWC   & SWC & SGN      & SGO    \\
    B & SGN   & RWC & SGO      & SWC    \\
    C & SWC   & SGO & RWC      & SGN    \\
    D & SGO   & SGN & SWC      & RWC    \\ \hline
  \end{tabular}
  \caption{Groups of evaluation experiments and order of task execution (order of execution from left to right).}\label{table:experiment-group}
\end{table}
\subsection{Results}\label{subsec10}
In the user study, 22 participants took part, all of whom were undergraduates or graduate students majoring in information science or had experience in the field.
Their ages ranged from 21 to 25, with an average of 21 years.
Nineteen participants were male, and three were female.
Sixteen were currently majoring in information visualization, while six had limited exposure to it.
None had prior experience with word cloud or other text visualization methods, though all were familiar with word cloud, and eight had never heard of semantic word cloud.

\begin{figure}[H]
  \centering
    \includegraphics[width=\linewidth]{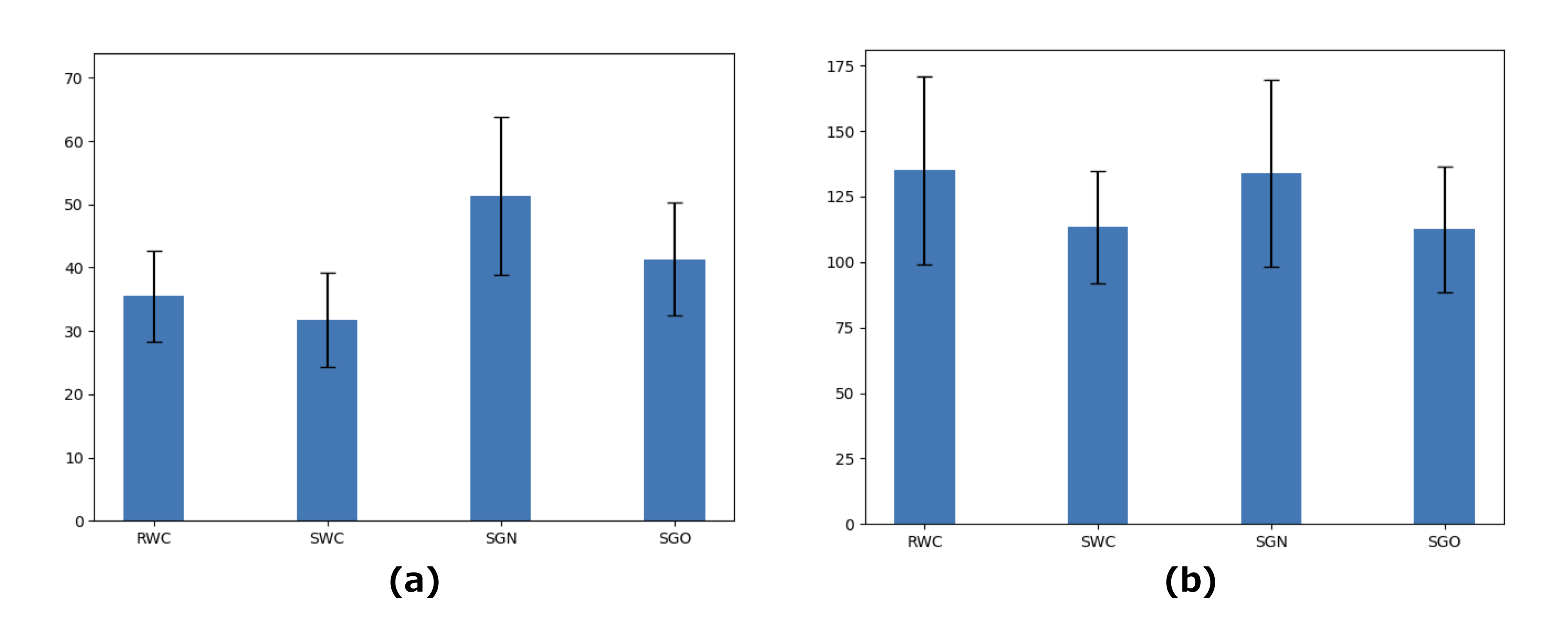}
  \caption{(a) Time (seconds) to perform a task of reading importance, and (b) Time (seconds) to perform a task of reading word relationships.}\label{fig:time-result}
\end{figure}

An analysis of variance examined differences between the four methods in the task execution time for importance and word relationship reading.
Significant differences were found in importance reading $(F(3, 22) = 3.6906, p=0.0150)$, though not in word relationship reading $(F(3, 22) = 1.2398, p=0.3004)$.
Fig. \ref{fig:time-result} summarizes the mean task execution times for the four methods, with error bars indicating 95\% confidence intervals.
The percentage of correct responses for each task was also assessed.
To perform a task of reading importance, participants were asked to identify the most and least frequent words, and then the percentage of correct responses was recorded for each word.
For the relationship reading task, participants were asked to identify words highly related to three specified words.
The five words most similar to those specified words, according to cosine similarity order, were considered the correct responses, and the percentage of correct responses was calculated accordingly.
Table \ref{table:tasks} presents the percentages of correct responses.
\begin{table}[H]
  \centering
  \begin{tabular}{|c|c|c|c|}
    \hline
        & max(N=22) & min(N=22) & similarity(N=66) \\ \hline
    RWC & 100\%     & 41\%      & 3\%              \\
    SWC & 95\%      & 50\%      & 9\%              \\
    SGN & 95\%      & 32\%      & 48\%             \\
    SGO & 95\%      & 60\%      & 50\%             \\ \hline
  \end{tabular}
  \caption{Percentage of correct responses to tasks associated with each visualization method. ``max'' and ``min'' columns show accuracy in identifying the most and least frequent words respectively. ``similarity'' column shows accuracy in identifying semantically related words.}\label{table:tasks}
\end{table}
They were also asked to answer which page of Wikipedia the applied data pertained to, and all participants were able to guess the original page for all methods of data.

\begin{figure}[H]
  \begin{center}
    \includegraphics[width=\linewidth]{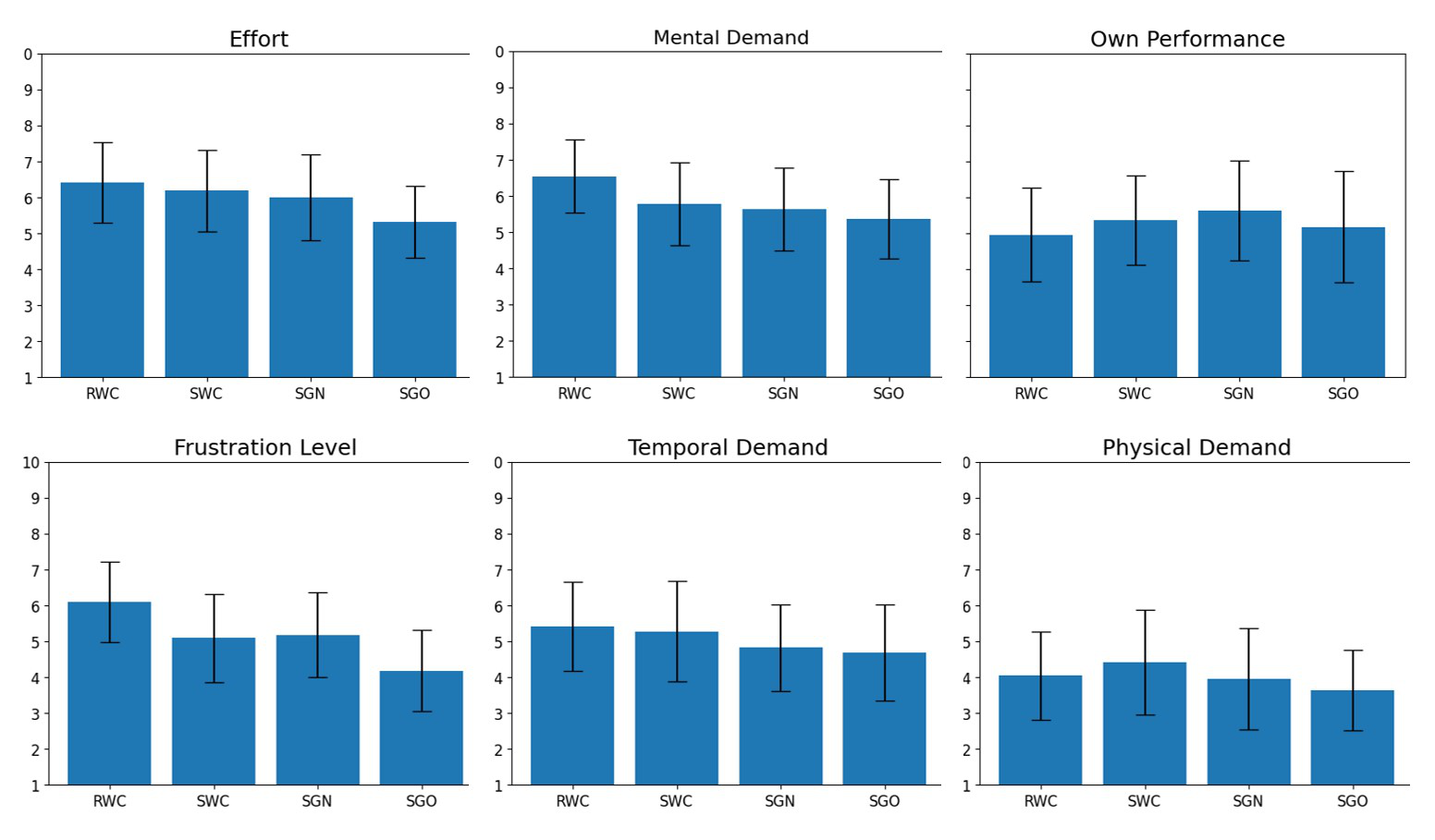}
    \caption{Results of the questionnaire based on NASA-TLX (lower is better).}\label{fig:nasa-tlx-results}
  \end{center}
\end{figure}

\begin{figure}[ht]
  \centering
    \includegraphics[width=0.7\linewidth]{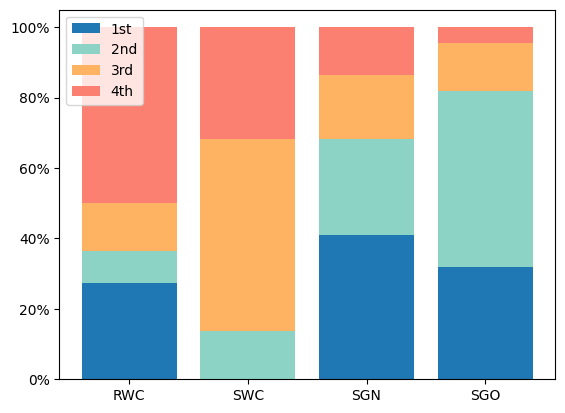}
  \caption{Percentage of visualization result preferences (N=22).}\label{fig:ranking}
\end{figure}

We analyzed the ratings of mental demand, physical demand, and task difficulty for performing tasks using four different visualization methods with the Friedman test and observed differences in the questionnaire scores.
The result of the Friedman test showed a significant difference in the frustration item $(\chi^2(3, N=22)=15.7499, p=0.0012)$.
Subsequently, Wilcoxon signed-rank tests were conducted, and the results showed that SGO scored significantly lower than RWC $(p=0.0111)$, SWC $(p=0.0143)$, and SGN $(p=0.0274)$.
Fig. \ref{fig:nasa-tlx-results} summarizes the results for the six NASA-TLX-based questions.
Bars indicate means, and error bars indicate 95\% confidence intervals.
Participants' subjective rankings of their preferences for the four visualization methods were analyzed with the Friedman test, and the difference in medians of the rankings of the four visualization methods was evaluated.
The result $(\chi^2(3, N=22)=15.2181, p=0.0016)$ showed a significant difference in the preference for the four visualization methods.
Subsequently, a Wilcoxon signed-rank test was conducted, and the results showed that SGO was significantly higher in rank than RWC $(p=0.0208)$ and SWC $(p=0.0001)$, and SGN was significantly higher in rank than SWC $(p=0.0006)$. There was no significant difference in ranking between RWC and SWC, and between SGN and SGO.
Fig. \ref{fig:ranking} summarizes the ranking percentages of which of the four visualization methods was the favorite among the four visualization methods.
Bars indicate means, and error bars indicate 95\% confidence intervals.
\subsection{User Feedback}\label{subsec11}
We asked participants to give their free opinions about their impressions of StoryGem compared to other text visualization methods.
The following is a summary of some of the actual positive feedback.
\begin{itemize}
  \item It is easy to understand what kind of content is visualized instantly, and the color coding reduces the effort to look at the content.
  \item The text is displayed in as large a font as possible, so it is easy to read.
  \item Important words were not missed.
  \item It was easy to read words that occur infrequently.
  \item It was easier to recognize letters if the font size was optimized.
\end{itemize}
Feedback confirmed that the intended visualization was achieved.
However, we also received some negative feedback below.
\begin{itemize}
  \item It was easy to read which words were related; however, it was difficult for me to find the closest element.
  \item I was bothered by the fact that the text was on the edge, perhaps to maximize the font size.
  \item StoryGem makes it difficult to see the relationship between different groups.
\end{itemize}
\subsection{Discussion}\label{subsec12}
Our experimental results revealed several key findings about StoryGem's performance compared to traditional visualization methods. \textbf{First}, in the importance reading task, we identified a significant limitation: StoryGem required more time for users to identify word frequencies compared to traditional approaches $(F(3, 22) = 3.6906, p=0.0150)$. This increased time requirement was particularly pronounced in the non-optimized version (SGN), which took significantly more time than both RWC $(p=0.0231)$ and SGO $(p=0.0174)$. While font size optimization (SGO) improved performance, the time required for importance reading remained higher than traditional methods, indicating a fundamental trade-off in our approach of using area size instead of font size to represent word importance. This limitation represents an important area for future improvement in our visualization method.

Despite this time-related disadvantage, we found that our approach offered better accuracy in identifying less frequent words (60\% for SGO vs. 41-50\% for other methods). This suggests that while our area-based importance representation may require more time to process, it provides better visibility and recognition accuracy for less prominent words, particularly when combined with font size optimization.

\textbf{Third}, and perhaps most significantly, StoryGem demonstrated remarkable effectiveness in conveying semantic relationships between words. In the word relationship reading task, while completion times were similar across all methods $(F(3, 22) = 1.2398, p=0.3004)$, StoryGem achieved dramatically higher accuracy rates (48-50\%) compared to traditional approaches (3-9\%). This five-fold improvement in accuracy without any time penalty represents a significant advancement in semantic relationship visualization.

\textbf{Fourth}, the mental workload evaluation through NASA-TLX revealed that StoryGem with optimization (SGO) significantly reduced user frustration compared to other methods (RWC: $p=0.0111$, SWC: $p=0.0143$, SGN: $p=0.0274$). This reduced frustration, combined with comparable mental and physical demands, suggests that our visualization approach provides a more comfortable user experience while maintaining effectiveness.

\textbf{Finally}, user preferences strongly favored StoryGem, with 80\% of participants ranking the optimized version (SGO) as their first or second choice. This preference was statistically significant compared to traditional visualizations (RWC: $p=0.0208$, SWC: $p=0.0001$), indicating that users not only found StoryGem effective but also enjoyable to use.

These findings suggest that while StoryGem's novel approach to representing word importance initially requires some adjustment from users, its benefits in semantic relationship visualization and overall user experience make it a valuable advancement in text visualization techniques. The optimization of font sizes proves crucial in bridging the gap between innovative design and practical usability.

\section{Conclusion}\label{sec6}
In this paper, we proposed a novel text visualization method called StoryGem, which arranges words within a predefined region while considering their semantic relationships.
By maximizing the font size within allocated areas, StoryGem improves text visibility in compact spaces and mitigates the influence of word length on visualization results.
To demonstrate the effectiveness of StoryGem, we conducted a user study and discussed the findings in detail.

Our results reveal several key advantages of StoryGem over traditional text visualization approaches. \textbf{First}, StoryGem significantly outperforms both random word clouds and semantic word clouds in conveying semantic relationships between words, with accuracy rates approximately five times higher (48-50\% vs. 3-9\%). \textbf{Second}, the font size optimization in StoryGem (SGO) leads to reduced frustration levels compared to other methods, creating a more pleasant user experience. \textbf{Third}, StoryGem with optimized font size was strongly preferred by users, with 80\% ranking it as their first or second choice among the four visualization methods.

These findings demonstrate that StoryGem successfully addresses the limitations of traditional word clouds (lack of semantic organization) and semantic word clouds (inefficient space utilization) by combining the strengths of both approaches. The use of Voronoi treemaps to create gapless layouts, coupled with our novel approach of representing word importance through region size rather than font size, provides a more balanced and accurate visual representation of textual data.

Despite these advantages, we acknowledge several limitations that present opportunities for future work. \textbf{First}, although StoryGem successfully represents semantic relationships through clustering, we have not yet achieved precise word positioning based on these relationships, as seen in semantic word clouds. Adjacent Voronoi regions within the same hierarchy are not guaranteed to represent closely related words, nor are differently colored adjacent regions necessarily semantically connected. This issue arises because the determination of Voronoi areas depends on the placement of initial seed points for centroidal Voronoi tessellation. We aim to address this limitation by optimizing the placement of these seed points based on word relationships, potentially incorporating force-directed algorithms to position semantically similar words closer together.

\textbf{Second}, some users reported difficulty in identifying relationships between different semantic groups in StoryGem. Future iterations could explore visual cues such as connecting lines or gradients between related clusters to better convey inter-group relationships.

\textbf{Third}, the current implementation places words at the edges of regions to maximize font size, which some users found distracting. We plan to explore alternative optimization approaches that balance maximizing font size with maintaining aesthetic placement within regions.

Additionally, we plan to enhance the system by developing a more intuitive user interface similar to popular word cloud generators, such as Wordle, where users can specify custom regions and see words arranged within these areas without gaps. We believe that StoryGem can be applied in various domains, including social media analysis, market research, literary studies, and educational settings, and we are committed to further refining and advancing this method to broaden its applicability and effectiveness.

\bibliography{reference}

\end{document}